# Accessible phases via wave impedance engineering with *PT*-symmetric metamaterials

Sotiris Droulias[1,2,*], Ioannis Katsantonis[1,2], Maria Kafesaki[1,2], Costas M. Soukoulis[1,3] and Eleftherios N. Economou[1,4]

[1]*Institute of Electronic Structure and Laser, Foundation for Research and Technology Hellas, 71110 Heraklion, Crete, Greece*
[2]*Department of Materials Science and Technology, University of Crete, 71003 Heraklion, Greece*
[3]*Ames Laboratory and Department of Physics and Astronomy, Iowa State University, Ames, Iowa 50011, USA*
[4]*Department of Physics, University of Crete, 71003 Heraklion, Greece*

Optical systems that respect Parity-Time (*PT*) symmetry can be realized with proper incorporation of gain/loss materials. However, due to the absence of magnetic response at optical frequencies, the wave impedance is defined entirely by their permittivity and, hence, the *PT*-symmetric character is controlled solely via their refractive index. Here, we show that the separate control of the wave impedance enabled by metamaterials can grant access to further tuning of the Exceptional Points, appearance of mixed phases (coexistence of *PT*-symmetric and *PT*-broken phases) and occurrence of phase re-entries, not easily realizable with natural materials.

Optical systems with gain and loss that respect Parity-Time (*PT*) symmetry can have real eigenvalues despite their non-Hermitian character; the eigenvalues remain real below some critical value of the potential, the so-called Exceptional Point (EP), above which they become complex and hence the EP marks the passing from the *PT*-symmetric phase to the broken-*PT* phase. This is an idea, originally introduced in the context of quantum mechanics [1-4], which quickly found fertile ground in optics due to the mathematical equivalence with paraxial beam propagation, which is described by a Schrödinger-like equation [5-11]. The extension of *PT*-symmetry to systems in which the eigenvalues refer to those of the scattering matrix [12-21] led to novel phenomena, such as coherent perfect absorption [13,14], the *PT*-laser absorber [15, 16], and anisotropic transmission resonances [18]. In such systems, the condition to achieve *PT*-symmetry is expressed in terms of the permittivity $\varepsilon$ and permeability $\mu$ as $\varepsilon(\mathbf{r}) = \varepsilon^*(-\mathbf{r})$ and $\mu(\mathbf{r}) = \mu^*(-\mathbf{r})$, where **r** is the position operator and the asterisk denotes the complex conjugate [22,23]. When realized with natural optical materials, the magnetic response is absent and, therefore, the *PT*-condition can be controlled only via $\varepsilon$. However, most recently, some works combined *PT*-symmetry with metamaterials [22-27], which could extend these ideas to new limits, as metamaterials can be designed to have the desired $\varepsilon$ and $\mu$, at almost any frequency [28].

While the ability to control the *PT*-phase can grant access to important properties, such as (a) mixed phases (coexistence of *PT*-symmetric and *PT*-broken phase) and (b) phase re-entries (multiple passes among all possible phases), these aspects have not been investigated thoroughly. For example, in [29] the occurrence of mixed phases was shown to require polarization converting elements, while phase re-entries were shown in [27] for TE waves in the special case of epsilon near zero (ENZ) metamaterials only.

In this work, we show that the coexistence of *PT*-symmetric and *PT*-broken phases occurs naturally in systems as simple as a one-dimensional gain/loss bilayer. These mixed phases emerge in oblique incidence [30-32] as a result of the different wave impedances of TE and TM linearly polarized waves, which are otherwise identical in normal incidence and therefore in that case the mixed phase vanishes. By properly engineering the wave impedance, we show that the passing of TE waves from the *PT*-symmetric to the *PT*-broken phase can precede, succeed or even coincide with that of the TM waves, thus allowing for tuning and eventually suppressing the mixed phase. We also show that, while natural materials favour a single Exceptional Point and thus a unique phase change, with metamaterials it is possible to engineer the wave impedance and observe multiple Exceptional Points and therefore phase re-entries. Last, an important aspect of our work is the formulation in terms of the refractive index *n* and the wave impedance $\zeta$. This approach provides a generalized description and deeper insight to the mechanism of phase change, as compared to all previous works where the analysis is based on $\varepsilon$ and $\mu$.

Considering one-dimensional systems, the conditions to achieve *PT*-symmetry for parameters changing e.g. along the *z*-direction are expressed in terms of the relative permittivity $\varepsilon$ and relative permeability $\mu$ as $\varepsilon(z) = \varepsilon^*(-z)$ and $\mu(z) = \mu^*(-z)$ [22,23]. These impose on the refractive index $n = \sqrt{\varepsilon\mu}$ and the wave impedance $\zeta = \sqrt{\mu/\varepsilon}$ to fulfill:

$$n(z) = n^*(-z), \quad \zeta(z) = \zeta^*(-z) \qquad (1)$$

Such conditions can be satisfied in the system of Fig. 1, which consists of two homogeneous gain/loss slabs which are infinite on the *xy*-plane and have finite length along the *z*-direction. Without loss of generality, gain (loss) is assumed to be embedded entirely in the left (right) slab.

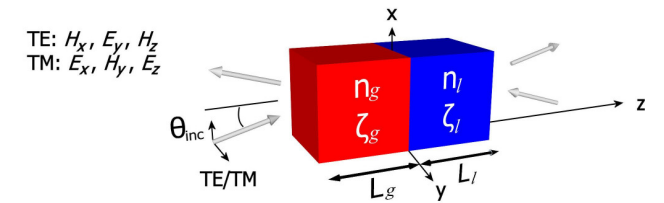

FIG. 1. A *PT*-symmetric heterostructure with a single gain/loss bilayer. The complex material parameters $n_i$, $\zeta_i$ are the refractive index and the wave impedance, respectively, and the subscript $i = \{g, l\}$ denotes whether they are located in the '*gain*' (red) or the '*loss*' (blue) region. For *PT*-symmetry the shown parameters satisfy $n_g = n_l^*$, $\zeta_g = \zeta_l^*$, and $L_g = L_l$ The incident and scattered waves can be either TE or TM polarized, as shown.

To study the scattering properties of the gain/loss bilayer, we assume that waves arrive at angle $\theta_{inc}$ from either side of the system (propagating along the *z*-direction) and we measure the scattered fields. The system is assumed to be in a homogeneous environment (air, for simplicity) and, hence, the waves exit the system at the same angle. Their polarization can be a mixture of TE components ($H_x$, $E_y$, $H_z$) and TM components ($E_x$, $H_y$, $E_z$) as shown in Fig.1. Although *PT*-symmetry requirements impose certain conditions in the loss and gain regions (see Eq. (1)), we start with slabs of arbitrary lengths, $L_g$, $L_l$ and arbitrary material parameters $n_g$, $n_l$, $\zeta_g$, $\zeta_l$ to obtain general expressions. Because the two polarization states are orthogonal to each other, the system can be described by two independent 2×2 scattering matrices, $S_{TE}$ and $S_{TM}$, corresponding to TE and TM waves, respectively (see Appendix for details). Each of the two matrices consists of two reflection and two transmission amplitudes, namely $r_L$, $r_R$, $t_L$, $t_R$, where the subscript *L*, *R* indicates incidence from '*Left*' or '*Right*', respectively. As in the case of normal incidence [16,18], we find for both polarizations that $r_L \neq r_R$ and $t_L = t_R \equiv t$. In general, $r_L$, $r_R$ and $t$ are different for TE and TM waves, except for normal incidence, where they become identical.

To identify whether the system lies in the *PT*-symmetric or *PT*-broken phase, we need to examine the eigenvalues $\lambda_{1,2}$ of *S* (*S* denoting $S_{TE}$ or $S_{TM}$), which have the general form $\lambda_{1,2} = (r_L + r_R \pm \sqrt{(r_L - r_R)^2 + 4t^2})/2$ [18]. Because $r_L$, $r_R$ and $t$ are different for TE and TM waves, the eigenvalues $\lambda_{1,2}$ are different as well for each of the two polarizations. For each individual $\lambda_{1,2}$ set, due to the presence of gain and loss, $|\lambda_{1,2}|\neq 1$ in general. However, if we apply *PT*-conditions an Exceptional Point emerges, below which $|\lambda_1|=|\lambda_2|=1$ (*PT*-symmetric phase), and $\lambda_1$, $\lambda_2$ become an inverse conjugate pair above it, satisfying $|\lambda_1||\lambda_2|=1$ [16,18] (*PT*-broken phase). In [16] it was shown that, for the eigenvalues of *S* to satisfy $|\lambda_1|=|\lambda_2|=1$ and therefore for the system to be in the *PT*-symmetric phase, the criterion is $|(r_L - r_R)/t|<2$. For each of the two polarizations, after some calculations we find that this condition can be written as:

$$\text{TE}: \left|\left(\frac{Z_l^{TE}}{Z_g^{TE}} - \frac{Z_g^{TE}}{Z_l^{TE}}\right)\sin(\delta_g)\sin(\delta_l)\right| < 2 \quad (2\text{a})$$

$$\text{TM}: \left|\left(\frac{Z_l^{TM}}{Z_g^{TM}} - \frac{Z_g^{TM}}{Z_l^{TM}}\right)\sin(\delta_g)\sin(\delta_l)\right| < 2 \quad (2\text{b})$$

where $Z_{g/l}^{TE} = \dfrac{\zeta_{g/l}}{\cos\theta_{g/l}}$, $Z_{g/l}^{TM} = \zeta_{g/l}\cos\theta_{g/l}$ is the wave impedance for TE and TM waves, respectively and $\delta_{g/l} = n_{g/l}(\omega/c) L_{g/l} \cos\theta_{g/l}$ ($n_{g/l}$, $\zeta_{g/l}$ are defined according to Eq.(1)). The angle $\theta_{g/l}$ is the wave propagation angle inside the *g*/*l* region and is defined by the continuity of the tangential *k*-components as $\sin(\theta_{inc}) = n_{g/l} \sin(\theta_{g/l})$ (the refractive index of the exterior –air– is unity).

With simple inspection, Eqs. (2a),(2b) have the same general form, which is a consequence of the duality of Maxwell's equations for E- and H- fields; however they are not identical. Due to the $\cos\theta_{g/l}$ term which appears asymmetrically in $Z_{g/l}^{TE}, Z_{g/l}^{TM}$, the term in parenthesis in the lhs of Eq.(2) is different among TE and TM waves and therefore *the Exceptional Points for the two polarizations are spontaneously different*. As a result, one polarization can pass to the *PT*-broken phase while the other still resides in the *PT*-symmetric phase, thus giving rise to a mixed *PT*-phase for waves of arbitrary polarization, such as unpolarized light. *The emergence of the mixed phase is thus a consequence of oblique incidence entirely*, as for normal incidence where $\theta_{inc} = 0$ and therefore $\theta_g = \theta_l = 0$, Eq.(2a),(2b) become identical and therefore the Exceptional Points of both polarizations coincide, as shown in Eq.(3):

$$\text{TE / TM}_{(\theta_{inc}=0)}: \left|\left(\frac{\zeta_l}{\zeta_g} - \frac{\zeta_g}{\zeta_l}\right)\sin\left(n_g\frac{\omega}{c}L_g\right)\sin\left(n_l\frac{\omega}{c}L_l\right)\right| < 2 \quad (3)$$

To demonstrate the above findings, we assume a nonmagnetic medium ($\mu_g = \mu_l = 1$) with $n_g = 2-0.2i$, $n_l = 2+0.2i$, as considered in previous works [18], and we scan the angle of incidence $\theta_{inc}$. The two slabs have equal length $L_g = L_l \equiv L/2$ and for each $\theta_{inc}$ we calculate the eigenvalues of $S_{TE}$, $S_{TM}$ as a function of the normalized frequency $\omega L/c$ (*c* is the vacuum speed of light). The cases where both polarizations are in the *PT*-symmetric or the *PT*-broken phase, are denoted in Fig. 2(a) as 'symmetric-*PT*' and 'broken-*PT*', respectively. The region marked as 'mixed-*PT*' denotes that one polarization has passed into the broken-*PT* phase, while the other still resides in the *PT*-symmetric phase. In Fig. 2(b) we show explicitly the calculated eigenvalues for $\theta_{inc} = 30$deg and $\theta_{inc} = 60$deg.

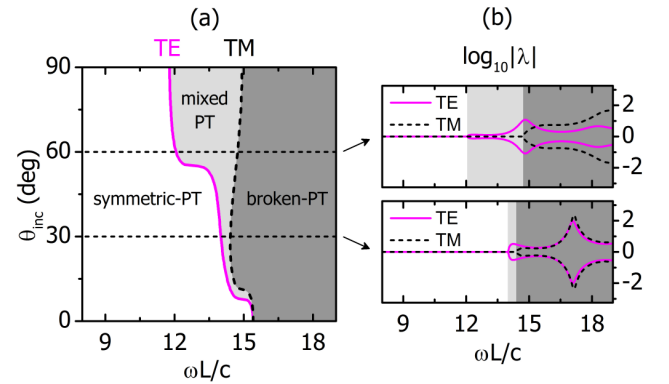

FIG. 2. Mixed phases as a consequence of oblique incidence, for a non-magnetic medium ($\mu = 1$) with $n_{g/l} = 2 \mp 0.2i$. (a) Phase diagram and (b) calculated scattering matrix eigenvalues for $\theta_{inc} = 30$deg and 60deg. In the 'symmetric-*PT*' and 'broken-*PT*' regions both TE and TM polarizations are in the same phase. In the 'mixed-*PT*' region TE waves have passed into the *PT*-broken phase, while TM waves still reside in the *PT*-symmetric phase.

Clearly, the Exceptional Points for TE and TM waves split spontaneously when we depart from normal incidence, but, if desired, the mixed phase can be suppressed even in oblique incidence, with properly engineering the wave impedance. This can be easily seen if we write the complex wave impedance for the gain and loss regions in polar form as $\zeta_g = |\zeta_g|exp(i\varphi_g)$ and $\zeta_l = |\zeta_l|exp(i\varphi_l)$, respectively. The *PT*-conditions impose $|\zeta_g|=|\zeta_l|\equiv|\zeta|$ and $\varphi_g = -\varphi_l \equiv \varphi$, i.e. $\zeta_g = |\zeta|e^{+i\varphi}$ and $\zeta_l = |\zeta|e^{-i\varphi}$. Then, the magnitude $|\zeta|$ is eliminated everywhere in Eqs. (2a),(2b) and terms of the form $e^{\pm 2i\varphi}$ appear, as shown in Eqs. (4a),(4b):

$$\text{TE}: \left|\left(e^{-2i\varphi}\frac{\cos\theta_g}{\cos\theta_l} - e^{+2i\varphi}\frac{\cos\theta_l}{\cos\theta_g}\right)\sin(\delta_g)\sin(\delta_l)\right| < 2 \quad (4a)$$

$$\text{TM}: \left|\left(e^{-2i\varphi}\frac{\cos\theta_l}{\cos\theta_g} - e^{+2i\varphi}\frac{\cos\theta_g}{\cos\theta_l}\right)\sin(\delta_g)\sin(\delta_l)\right| < 2 \quad (4b)$$

This result implies that the positions of the Exceptional Points are not expected to depend on the magnitude of $\zeta$, but solely on the relative strength between its real and imaginary part, which is expressed via $\varphi$. Simple observation leads to the conclusion that Eqs. (4a),(4b) become identical if $\varphi$ becomes multiples of $\pi/4$. This means that the Exceptional Points of TE and TM waves coincide if $|\text{Re}(\zeta_{g/l})|=|\text{Im}(\zeta_{g/l})|$ or $|\text{Re}(\zeta_{g/l})| = 0$ or $|\text{Im}(\zeta_{g/l})| = 0$ and this causes the mixed phases to vanish. Furthermore, a sign flip in $\varphi$ interchanges Eqs. 4(a),(b), i.e. the Exceptional Points of TE and TM waves exchange positions. From this analysis it is also evident that the refractive index does not participate in the tailoring of the mixed phases. This is not surprising, as $n$ appears only in the arguments of the sine terms in Eqs.(4a),(4b), which are identical for both expressions and, hence, *the mixed phases are tailored via the wave impedance entirely*. We note here that with natural gain/loss materials in which $\mu_g = \mu_l = 1 \rightarrow \zeta_{g/l} = (1/\varepsilon_{g/l})^{1/2} = 1/n_{g/l}$, and therefore $|\text{Re}(\zeta_{g/l})|/|\text{Im}(\zeta_{g/l})| = |\text{Re}(n_{g/l})|/|\text{Im}(n_{g/l})|$. Due to this result, because $|\text{Re}(n_{g/l})|>>|\text{Im}(n_{g/l})|$ it follows that $|\text{Re}(\zeta_{g/l})|\neq|\text{Im}(\zeta_{g/l})|$ and, hence, the mixed phases appear naturally; to eliminate them, independent tuning of the wave impedance is required, i.e. a magnetic response is necessary.

To demonstrate the above findings, we return to the system of the previous example with $n_g = 2-0.2i$, $n_l = 2+0.2i$, for which we now allow for magnetic response. We set $|\zeta| = 0.5$ (~ $|1/n_{g/l}|$) and tune the wave impedance via the angle $\varphi$. Figure 3 shows examples for $\varphi = +3$ deg. (panel (a)), $\varphi = 0$ (panel (b)) and $\varphi = -3$ deg. (panel (c)). As predicted, the phase separation does not depend on the magnitude of $\zeta$, but solely on $\varphi$ ($|\zeta|$ just adjusts the magnitude of the eigenvalues in the broken-*PT* phase and not the position of the Exceptional Point, see Appendix). A sign flip in $\varphi$ exchanges the position of the TE and TM Exceptional Points [compare Fig.3(a),(c)] and for $\varphi = m\times\pi/4$ (*m*: integer) the mixed phase is completely suppressed as shown in (b), which corresponds to the case of $m = 0$.

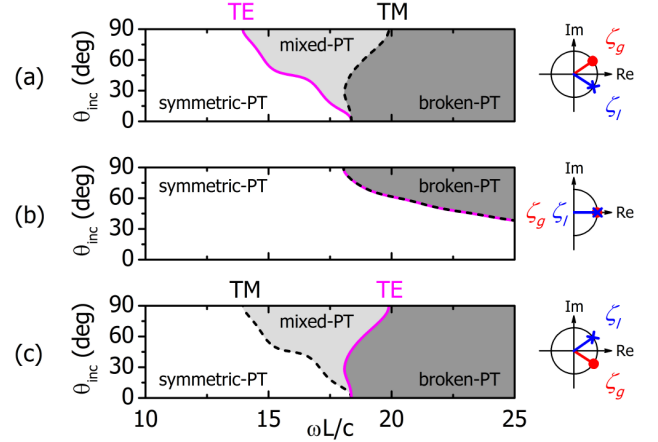

FIG. 3. Engineering of mixed phases via the wave impedance, $\zeta_{g/l} = |\zeta|e^{\pm i\varphi}$, for a system with $n_{g/l} = 2 \mp 0.2i$ and $|\zeta| = 0.5$. (a) $\varphi = +3$ deg. (b) $\varphi = 0$ (c) $\varphi = -3$ deg. The phase separation depends solely on $\varphi$ (not on $|\zeta|$, which just adjusts the magnitude of the eigenvalues in the broken-*PT* phase). A sign flip in $\varphi$ exchanges the position of the TE and TM Exceptional Points [compare (a), (c)] and for $\varphi = m\times\pi/4$ (*m*: integer) the mixed phase is completely suppressed, as shown in (b) for $m = 0$.

Such extraordinary cases, which are difficult to observe with natural non-magnetic optical materials, can be accomplished with metamaterials, which allow for tailored electric and magnetic response, via their resonances [28]. Engineering the wave impedance with metamaterials can also grant access to phenomena not easily obtainable with natural materials, such as phase re-entries, as a consequence of multiple Exceptional Points. To understand how this is possible, let us consider the case of normal incidence, for which Eq. (3) simplifies considerably and helps elucidate the important aspects. If we write the *PT*-symmetric $n$ and $\zeta$ as $n_{g/l} = n' \mp in''$, $\zeta_{g/l} = \zeta' \pm i\zeta''$, Eq.(3) is now expressed in terms of the real quantities $n'$, $n''$, $\zeta'$, $\zeta''$ as:

$$\text{TE / TM} \atop (\theta_{inc}=0) : \left|\frac{2i\zeta'\zeta''}{\zeta'^2 + \zeta''^2}\left(\cos\left(n'\frac{\omega L}{c}\right) - \cosh\left(n''\frac{\omega L}{c}\right)\right)\right| < 2 \quad (5)$$

The lhs of this inequality consists of an oscillatory part (cos–term) with amplitude $2\zeta'\zeta''/(\zeta'^2+\zeta''^2)$, which undergoes an exponentially growing offset (cosh–term) with a rate that depends on the strength of $n''$, i.e. $\zeta$ tunes the oscillation amplitude, while $n$ tunes its offset. Phase re-entries means that the lhs of Eq. (5) exceeds the rhs multiple times as function of $\omega L/c$. This requires strong oscillatory amplitude, which is maximized for $\zeta' = \pm\zeta''$, along with suspended offset, i.e. relatively weak $n''$. We note here that with natural gain/loss materials in which $\zeta_{g/l} = 1/n_{g/l}$, the oscillation amplitude becomes $2n'n''/(n'^2+n''^2)$ and, hence, the need for strong oscillation amplitude is now expressed as $n''=n'$, thereby contradicting the need for weak amplitude offset, i.e. $n''<<n'$. Because both requirements involve $n''$ in the opposite manner, it is somewhat difficult to achieve phase re-entries with natural materials; however, with metamaterials, such cases are possible.

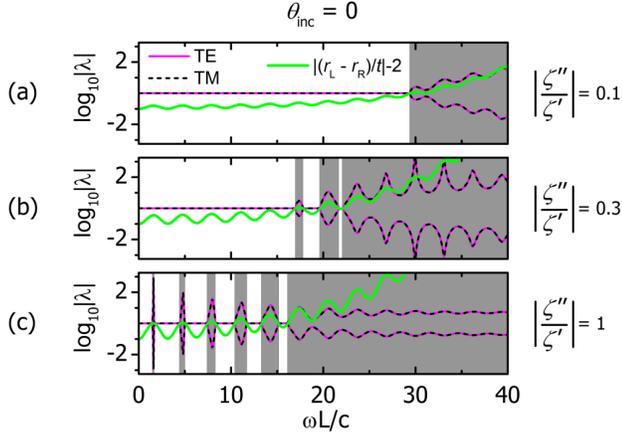

FIG. 4. Explanation of phase re-entries via wave impedance engineering for a system with $n_{g/l} = 2 \mp 0.1i$ and $\zeta_{g/l} = \zeta' \pm i\zeta''$ (normal incidence). The ratio $|\zeta''/\zeta'|$ tunes the oscillation amplitude of the lhs of Eq.(5), plotted as green line. (a) Weak oscillation for $|\zeta''/\zeta'| = 0.1$ and single Exceptional Point. (b) Intermediate oscillation for $|\zeta''/\zeta'| = 0.3$ and initiation of phase re-entries. (c) Strong oscillation for $|\zeta''/\zeta'| = 1$, leading to multiple phase re-entries. For oblique incidence these results become richer, as mixed phases are introduced.

In Fig. 4 we show the eigenvalues for a system with $n' = 2$ and $n'' = 0.1$ (i.e. $n_{g/l} = 2 \mp 0.1i$) in normal incidence, for which $\zeta' = 1$, and $\zeta''$ is tuned among three distinct values, namely 0.1, 0.3 and 1, in order to adjust the oscillation amplitude of the lhs of Eq. (5), i.e. $|(r_L - r_R)/t|$ (plotted here as $|(r_L - r_R)/t|-2$ and shown as a green line). For $\zeta'' = 0.1$ (weak oscillation) the system passes on to the broken-PT phase at $\omega L/c = 29.4$. As $\zeta''$ becomes stronger, the oscillation becomes stronger and for $\zeta'' = 0.3$ a double phase re-entry occurs in the region $\omega L/c \sim 17-22$. Last, for $\zeta'' = 1$ we observe several passes from the PT-symmetric to the broken-PT phase in the region $\omega L/c \sim 0-16$. At the same time, the Exceptional Points are shifted to lower $\omega L/c$, illustrating how their position can be further tailored via wave impedance engineering. The latter case ($\zeta_{g/l} = 1 \pm i$) maximizes the oscillation amplitude in the lhs of Eq. (5), because it corresponds to $\zeta' = \zeta''$, and therefore provides the maximum re-entry effect. Additionally, $\zeta' = \zeta''$ corresponds to $\varphi = \pm\pi/4$ in accordance with our previous analysis and therefore the mixed phase is completely eliminated when we examine the same system in oblique incidence. In general, though, there is no need for such strict condition and both re-entry effects and mixed phases can be still achieved if $|\zeta'|$, $|\zeta''|$ are not equal, especially at large angles where the $\cos\theta_{g/l}$ terms become significant. This will become apparent in the following example. In any case, the important conclusion is that, for a fixed refractive index $n_{g/l}$, *all tunability depends on the ratio $|\zeta''/\zeta'|$ and not on the magnitude of $\zeta$*; the same phase diagrams correspond to any wave impedance with the same relative strength between its real and imaginary part. In other words, these graphs show a parametric family, which can be achieved with several different sets of $\varepsilon$ and $\mu$.

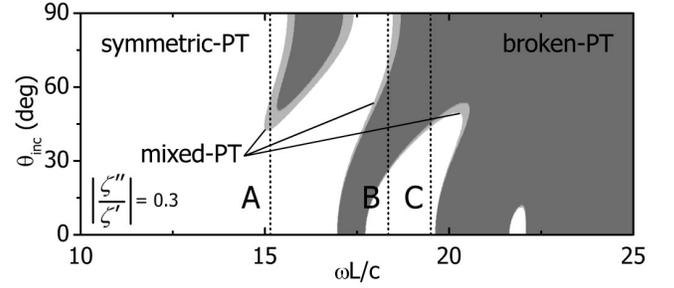

FIG. 5. Universal phase diagram for a system with $n = 2 \mp 0.1i$ and $|\zeta''/\zeta'| = 0.3$, corresponding to several choices for $\varepsilon$ and $\mu$. The possibilities for (a) mixed phases and (b) phase re-entries can be achieved as function of either $\omega L/c$ or $\theta_{inc}$, as denoted with the dashed vertical lines. The marked cases show partial phase re-entry (line A), full phase re-entry (line B) and typical phase change from PT-symmetric to the PT-broken phase (line C). Note that the cross-section at $\theta_{inc} = 0$ corresponds to the plot shown in Fig. 4(b).

The results shown so far in terms of the normalized parameter $\omega L/c$ imply constant material parameters, which are not easily realizable with real, dispersive materials [21]. In practice, however, because the optical potential involves both the refractive index and the wave impedance –besides the frequency–, the PT-transition can be observed at a single frequency, with varying the values of gain and loss or just the incidence angle $\theta_{inc}$ (which tunes the angle $\theta_{g/l}$ that appears in $Z_{g/l}^{TE}, Z_{g/l}^{TM}$). To demonstrate this alternative, in Fig. 5 we show the phase diagram of the previous system with $n_g = 2-0.1i$, $n_l = 2+0.1i$ and $|\zeta''/\zeta'| = 0.3$ for $\theta_{inc} \neq 0$, i.e. we expand the case shown for normal incidence in Fig. 4(b) to account for oblique incidence. This reveals a rich behavior and the possibilities for mixed phases and phase re-entries as function of either $\omega L/c$ or $\theta_{inc}$. The vertical dashed lines demonstrate three characteristic cases, which are accessible with scanning $\theta_{inc}$, while keeping all the other parameters fixed. From this example it is evident that partial phase re-entries (line A), full phase re-entries (line B) and typical phase changes (line C) are possible all within the same system with constant material parameters. Note also that in the examples shown in Fig. 4 and Fig. 5 we have assumed that $|\zeta''/\zeta'| \leq 1$. When the values of $\zeta''$, $\zeta'$ are interchanged, this ratio is reversed, the TE and TM Exceptional Points exchange positions and therefore the phase diagram remains the same (see Appendix for details).

In conclusion, we have shown that the coexistence of PT-symmetric, PT-broken, and mixed phases is possible even in simple one-dimensional photonic heterostructures with a single gain/loss bilayer. This plethora of phases including re-entry behaviours (see Fig. 5) emerge as a result of the different wave impedances, properly engineered, between TE and TM linearly polarized waves; by varying the relative strength of these impedances simply through the *angle of oblique incidence* the obtained very rich phase diagram is achieved without requiring polarization converting elements or other additional components. We have further shown that the Exceptional Points of TE and TM waves can be tuned to modify and even suppress the mixed phase. We have also shown that, while natural materials favour a single Exceptional Point and thus a unique passing from *PT*-symmetric to *PT*-broken phase, with metamaterials it is possible to engineer the wave impedance (independently of the refractive index) to observe multiple Exceptional Points and therefore phase re-entries. All the above possibilities become clear when the system is examined under the prism of $n$ and $\zeta$, rather than $\varepsilon$ and $\mu$. Specific cases examined in previous works all fall within our generalized approach, which provides a unified description, deep insight and specific guidelines for the design of the desired response.

## Acknowledgements

This work was supported by the Hellenic Foundation for Research and Innovation (HFRI) and the General Secretariat for Research and Technology (GSRT), under the HFRI PhD Fellowship grant (GA. no. 4820). It was also supported by the EU-Horizon2020 FET projects Ultrachiral and Visorsurf.

## Appendix

### 1. Scattering coefficients for TE and TM waves (oblique incidence)

To find the reflection and transmission amplitudes of the double-slab system we solve Maxwell's equations with the boundary conditions at each material interface. We assume that waves arrive at angle $\theta_{inc}$ from either side of the system, with polarization which can be a mixture of TE components ($H_x$, $E_y$, $H_z$) and TM components ($E_x$, $H_y$, $E_z$) as shown in Fig.1. For simplicity we assume that the surrounding space is air and, therefore the wavenumber outside the double slab is $k_0$, the free-space wavenumber. Inside each slab region the waves propagate at angle $\theta_i$, (the subscript $i = \{g, l\}$ denotes the '*gain*' and '*loss*' region, respectively) and is given by $k_0 \sin(\theta_{inc}) = k_g \sin(\theta_g) = k_l \sin(\theta_l)$, which results from the continuity of the tangential field components at each interface. In the last relation $k_{g/l}$ is the wavenumber in each region, which is given by $k_{g/l} = k_0 \sqrt{\varepsilon_{g/l} \mu_{g/l}} \equiv k_0 n_{g/l}$ thereby reducing the relation to $\sin(\theta_{inc}) = n_{g/l} \sin(\theta_{g/l})$, as presented in the main text. To satisfy the *PT*-symmetry requirements given by Eq. (1), we are interested in material parameters of certain spatial symmetry; however, we start with slabs of arbitrary properties, $\varepsilon_i$, $\mu_i$ and $L_i$, to obtain general expressions. The general (non-*PT*) analytical expressions for TE and TM incident waves are listed below. The letter *L/R* in the subscript denotes incidence from '*Left*'/'*Right*'.

**TE polarization ($H_x$, $E_y$, $H_z$):**

$$r_L^{TE} = -e^{-2iL_g k_0 \cos\theta_{inc}} \frac{C_L^{TE} + D_L^{TE}}{A^{TE} + B^{TE}}$$

$$r_R^{TE} = -e^{-2iL_l k_0 \cos\theta_{inc}} \frac{C_R^{TE} + D_R^{TE}}{A^{TE} + B^{TE}}$$

and $t_L^{TE} = t_R^{TE} \equiv t^{TE}$ with

$$t^{TE} = -8e^{-ik_0(L_g + L_l)\cos\theta_{inc}} e^{i(L_g k_g \cos\theta_g + L_l k_l \cos\theta_l)} \frac{R_g R_l \cos\theta_{inc}}{A^{TE} + B^{TE}}$$

where $R_g = \dfrac{\cos\theta_g}{\cos\theta_{inc}}$, $R_l = \dfrac{\cos\theta_l}{\cos\theta_{inc}}$ and:

**TM polarization ($E_x$, $H_y$, $E_z$):**

$$r_L^{TM} = -e^{-2iL_g k_0 \cos\theta_{inc}} \frac{C_L^{TM} + D_L^{TM}}{A^{TM} + B^{TM}} \qquad (A1)$$

$$r_R^{TM} = -e^{-2iL_l k_0 \cos\theta_{inc}} \frac{C_R^{TM} + D_R^{TM}}{A^{TM} + B^{TM}}$$

and $t_L^{TM} = t_R^{TM} \equiv t^{TM}$ with

$$t^{TM} = -8e^{-ik_0(L_g + L_l)\cos\theta_{inc}} e^{i(L_g k_g \cos\theta_g + L_l k_l \cos\theta_l)} \frac{R_g R_l \cos\theta_{inc}}{A^{TM} + B^{TM}}$$

$$A^{TE} = Z_g \cos\theta_l \left( +\left(\frac{R_g}{Z_g}+1\right) + \left(\frac{R_g}{Z_g}-1\right)e^{2iL_g k_g \cos\theta_g} \right)\left( -\left(\frac{R_l}{Z_l}+1\right) + \left(\frac{R_l}{Z_l}-1\right)e^{2iL_l k_l \cos\theta_l} \right)$$

$$B^{TE} = Z_l \cos\theta_g \left( -\left(\frac{R_g}{Z_g}+1\right) + \left(\frac{R_g}{Z_g}-1\right)e^{2iL_g k_g \cos\theta_g} \right)\left( +\left(\frac{R_l}{Z_l}+1\right) + \left(\frac{R_l}{Z_l}-1\right)e^{2iL_l k_l \cos\theta_l} \right)$$

$$C_L^{TE} = Z_g \cos\theta_l \left( +\left(\frac{R_g}{Z_g}+1\right)e^{2iL_g k_g \cos\theta_g} + \left(\frac{R_g}{Z_g}-1\right) \right)\left( +\left(\frac{R_l}{Z_l}-1\right)e^{2iL_l k_l \cos\theta_l} - \left(\frac{R_l}{Z_l}+1\right) \right)$$

$$D_L^{TE} = Z_l \cos\theta_g \left( +\left(\frac{R_g}{Z_g}+1\right)e^{2iL_g k_g \cos\theta_g} - \left(\frac{R_g}{Z_g}-1\right) \right)\left( +\left(\frac{R_l}{Z_l}-1\right)e^{2iL_l k_l \cos\theta_l} + \left(\frac{R_l}{Z_l}+1\right) \right)$$

$$C_R^{TE} = Z_g \cos\theta_l \left( +\left(\frac{R_g}{Z_g}-1\right)e^{2iL_g k_g \cos\theta_g} + \left(\frac{R_g}{Z_g}+1\right) \right)\left( +\left(\frac{R_l}{Z_l}+1\right)e^{2iL_l k_l \cos\theta_l} - \left(\frac{R_l}{Z_l}-1\right) \right)$$

$$D_R^{TE} = Z_l \cos\theta_g \left( +\left(\frac{R_g}{Z_g}-1\right)e^{2iL_g k_g \cos\theta_g} - \left(\frac{R_g}{Z_g}+1\right) \right)\left( +\left(\frac{R_l}{Z_l}+1\right)e^{2iL_l k_l \cos\theta_l} + \left(\frac{R_l}{Z_l}-1\right) \right)$$

$$A^{TM} = Z_g \cos\theta_l \left( +\left(\frac{1}{Z_g}+R_g\right) + \left(\frac{1}{Z_g}-R_g\right)e^{2iL_g k_g \cos\theta_g} \right)\left( -\left(\frac{1}{Z_l}+R_l\right) + \left(\frac{1}{Z_l}-R_l\right)e^{2iL_l k_l \cos\theta_l} \right)$$

$$B^{TM} = Z_l \cos\theta_l \left( -\left(\frac{1}{Z_g}+R_g\right) + \left(\frac{1}{Z_g}-R_g\right)e^{2iL_g k_g \cos\theta_g} \right)\left( +\left(\frac{1}{Z_l}+R_l\right) + \left(\frac{1}{Z_l}-R_l\right)e^{2iL_l k_l \cos\theta_l} \right)$$

$$C_L^{TM} = Z_g \cos\theta_l \left( +\left(\frac{1}{Z_g}+R_g\right)e^{2iL_g k_g \cos\theta_g} + \left(\frac{1}{Z_g}-R_g\right) \right)\left( +\left(\frac{1}{Z_l}-R_l\right)e^{2iL_l k_l \cos\theta_l} - \left(\frac{1}{Z_l}+R_l\right) \right)$$

$$D_L^{TM} = Z_l \cos\theta_l \left( +\left(\frac{1}{Z_g}+R_g\right)e^{2iL_g k_g \cos\theta_g} - \left(\frac{1}{Z_g}-R_g\right) \right)\left( +\left(\frac{1}{Z_l}-R_l\right)e^{2iL_l k_l \cos\theta_l} + \left(\frac{1}{Z_l}+R_l\right) \right)$$

$$C_R^{TM} = Z_g \cos\theta_l \left( +\left(\frac{1}{Z_g}-R_g\right)e^{2iL_g k_g \cos\theta_g} + \left(\frac{1}{Z_g}+R_g\right) \right)\left( +\left(\frac{1}{Z_l}+R_l\right)e^{2iL_l k_l \cos\theta_l} - \left(\frac{1}{Z_l}-R_l\right) \right)$$

$$D_R^{TM} = Z_l \cos\theta_l \left( +\left(\frac{1}{Z_g}-R_g\right)e^{2iL_g k_g \cos\theta_g} - \left(\frac{1}{Z_g}+R_g\right) \right)\left( +\left(\frac{1}{Z_l}+R_l\right)e^{2iL_l k_l \cos\theta_l} + \left(\frac{1}{Z_l}-R_l\right) \right)$$

The parameter $Z_i = \sqrt{\mu_i/\varepsilon_i}$, $i = \{g,l\}$, is the wave impedance, which is normalized to the free-space impedance $Z_0 = \sqrt{\mu_0/\varepsilon_0}$. For normal incidence, i.e. $\theta_{inc} = 0 \rightarrow \theta_g = \theta_l = 0$ and $R_g = R_l = 1$, $\cos\theta_g = \cos\theta_l = 1$. With simple substitution it is easy to verify that the scattering coefficients of the two polarizations become identical.

## 2. Scattering matrix and eigenvalues for TE and TM waves

Because the two polarization states are orthogonal to each other, the system can be examined separately for each polarization. The scattering matrix $S$ for TE (TM) waves is denoted as $S_{TE}$ ($S_{TM}$) and consists of two reflection and two transmission coefficients, namely $r_L$, $r_R$, $t_L$, $t_R$. Following the standard definition [16,18], $S$ is written explicitly for each polarization as:

$$S^{TE} = \begin{bmatrix} r_L^{TE} & t_R^{TE} \\ t_L^{TE} & r_R^{TE} \end{bmatrix} \equiv \begin{bmatrix} r_L^{TE} & t^{TE} \\ t^{TE} & r_R^{TE} \end{bmatrix} \text{ and } S^{TM} = \begin{bmatrix} r_L^{TM} & t_R^{TM} \\ t_L^{TM} & r_R^{TM} \end{bmatrix} \equiv \begin{bmatrix} r_L^{TM} & t^{TM} \\ t^{TM} & r_R^{TM} \end{bmatrix} \quad (A2)$$

Both $S$-matrices satisfy the condition $PT\,S(\omega^*)\,PT = S^{-1}(\omega)$ [16,18]. The corresponding eigenvalues are:

$$\lambda_{1,2}^{TE} = \frac{1}{2}\left( r_L^{TE} + r_R^{TE} \pm \sqrt{\left(r_L^{TE} - r_R^{TE}\right)^2 + 4\left(t^{TE}\right)^2} \right) \text{ and } \lambda_{1,2}^{TM} = \frac{1}{2}\left( r_L^{TM} + r_R^{TM} \pm \sqrt{\left(r_L^{TM} - r_R^{TM}\right)^2 + 4\left(t^{TM}\right)^2} \right) \quad (A3)$$

## 3. Condition for PT-symmetric phase and further examples

Previously, in order to clarify that the positions of the Exceptional Points for TE and TM waves do not depend on the magnitude of the wave impedance, but solely on the relative strength between its real and imaginary part, we expressed $\zeta_g$ and $\zeta_l$ in polar form. If, instead, we write the PT-symmetric $\zeta$ as $\zeta_{g/l} = \zeta' \pm i\zeta''$, ($\zeta'$, $\zeta''$ : real quantities) then the criterion $|(r_L - r_R)/t|<2$ for residing in the PT-symmetric phase as expressed in Eq. (2a),(2b) now takes the form:

TE: $\left| \left( \frac{\zeta' - i\zeta''}{\zeta' + i\zeta''}\frac{\cos\theta_g}{\cos\theta_l} - \frac{\zeta' + i\zeta''}{\zeta' - i\zeta''}\frac{\cos\theta_l}{\cos\theta_g} \right) \sin(\delta_g)\sin(\delta_l) \right| < 2$

TM: $\left| \left( \frac{\zeta' - i\zeta''}{\zeta' + i\zeta''}\frac{\cos\theta_l}{\cos\theta_g} - \frac{\zeta' + i\zeta''}{\zeta' - i\zeta''}\frac{\cos\theta_g}{\cos\theta_l} \right) \sin(\delta_g)\sin(\delta_l) \right| < 2$

(A4)

In this form it is easy to observe that, depending on the sign of $\zeta'$ and $\zeta''$, the positions of the Exceptional Points either exchange between TE and TM waves or do not change at all. Simply put, the range of the PT-symmetric, mixed and broken-PT phase remain the same regardless of the exact sign of $\zeta'$ and $\zeta''$, which only defines which of the two polarizations crosses the EP first. Hence, all sign combinations for $\zeta'$, $\zeta''$ that yield the same $|\zeta''/\zeta'|$ ratio fall under the same universal phase diagram. Additionally, with exchanging $\zeta' \leftrightarrow \zeta''$ the phase diagram is preserved, because we can restore the initial Eqs. (A4) by multiplying each wave impedance $\zeta$ by $i$. Hence, under the above transformations the positions of the Exceptional Points are immobile and a certain phase diagram may correspond to more than one wave impedances; however the magnitude of the eigenvalues $\lambda$ changes in general within each broken-PT phase (in the symmetric-PT phase all eigenvalues satisfy $|\lambda| = 1$).

To demonstrate these conclusions, in Fig. A1 we return to the system of Fig. 4, 5 with $n_g = 2-0.1i$ and $n_l = 2+0.1i$ and examine the cases with (a) $\zeta' = 1$, $\zeta'' = +0.1$, (b) $\zeta' = 1$, $\zeta'' = -0.1$, (c) $\zeta' = 0.1$, $\zeta'' = +1$ and (d) $\zeta' = 0.1$, $\zeta'' = -1$. These choices correspond to either $|\zeta''/\zeta'| = 0.1$ or $|\zeta'/\zeta''| = 0.1$ and provide therefore the same universal phase diagram, which is shown in Fig. A1(a). The eigenvalues for each case are shown in Fig. A1(b)-(e) for $\theta_{inc} = 60$deg, as denoted with the dashed line in Fig. A1(a). Note that in all cases the magnitude of $\zeta_{g/l}$ is the same. Indeed, as we mentioned in the manuscript, the positions of the Exceptional Points do not depend on the magnitude of $\zeta$, but solely on the relative strength between its real and imaginary part.

Last, we show how the mixed phases can be tuned with changing the relative strength of $\zeta'$, $\zeta''$. In Fig. A2 we show the phase diagrams as the ratio $|\zeta''/\zeta'|$ increases from 0.2 to 1. The mixed phase is being gradually suppressed until $|\zeta''/\zeta'| = 1$, where it is completely eliminated. According to our previous analysis in terms of $\zeta_g = |\zeta|e^{+i\varphi}$ and $\zeta_l = |\zeta|e^{-i\varphi}$, this is a case where $\varphi$ is multiple of $\pi/4$, leading the Exceptional Points of TE and TM waves to coincide.

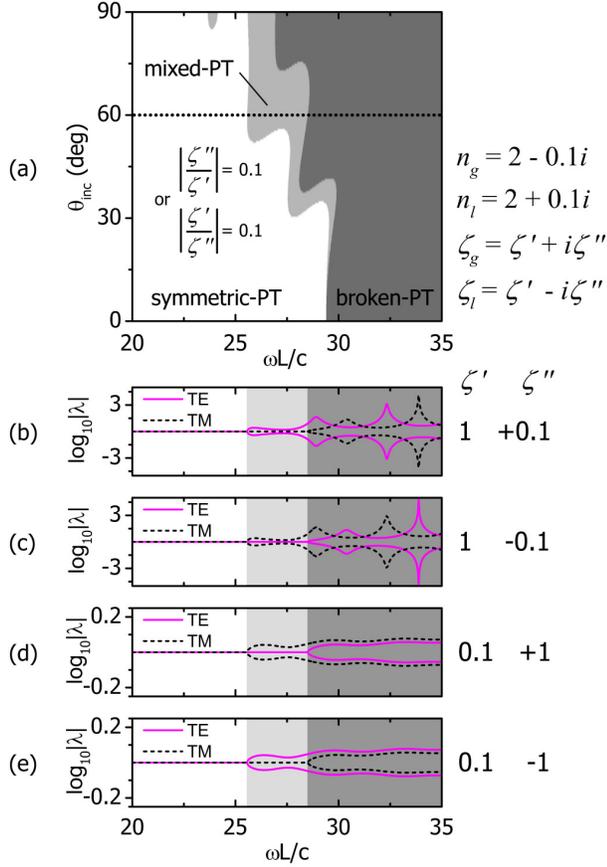
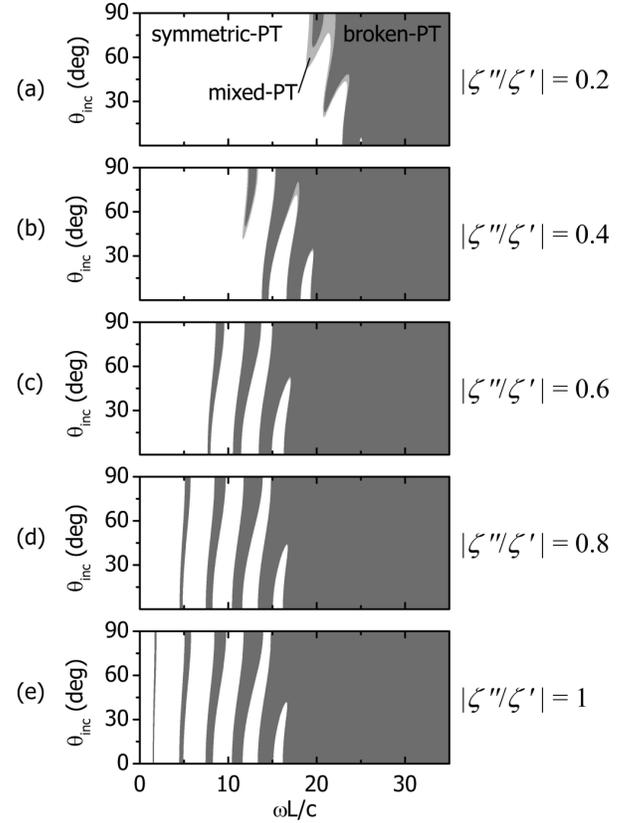

FIG. A.1. Universal phase diagram and interchangeability of properties between TE and TM waves. (a) Universal phase diagram of the system of Fig.4 with $n_{g/l} = 2 \mp 0.1i$ and $\zeta_{g/l} = \zeta' \pm i\zeta''$ with $\zeta' =1$, $\zeta'' = \pm 0.1$ or $\zeta' = 0.1$, $\zeta'' = \pm 1$. Under these choices the range of the symmetric-$PT$, mixed and broken-$PT$ phase is preserved. The eigenvalues vary upon the individual choice for $\zeta'$, $\zeta''$ and are shown for $\theta_{inc} = 60$deg and choice of (b) $\zeta' = 1$, $\zeta'' = +0.1$, (c) $\zeta' = 1$, $\zeta'' = -0.1$, (d) $\zeta' = 0.1$, $\zeta'' = +1$ and (e) $\zeta' = 0.1$, $\zeta'' = -1$. All choices share the same phase diagram, however, the Exceptional Points of TE and TM waves may interchange positions. This general property owes its validity to the symmetric way that the wave impedances of TE and TM waves appear in Eq. (2).

FIG. A.2. Universal phase diagram and phase transformations. The system of Fig.5 with $n = 2 \mp 0.1i$ is examined here, where the ratio $|\zeta''/\zeta'|$ is varied within the 0-1 range. Intermediate stages for $|\zeta''/\zeta'|$ = 0.2, 0.4, 0.6, 0.8 and 1 are shown to demonstrate the phase transformations. With increasing $|\zeta''/\zeta'|$ the mixed phase is being gradually suppressed until $|\zeta''/\zeta'| = 1$, where it is completely eliminated.